\documentclass[a4paper,11pt]{article}

\usepackage{orcidlink}
\usepackage{authblk}

\usepackage[backend=biber,     
	bibstyle=ieee,              
	citestyle=numeric-comp,     
	sortcites=true,
	maxbibnames=3,
	bibencoding=utf8]{biblatex}
\usepackage{hyperref}

\usepackage[acronyms,postdot]{glossaries-extra}

\newignoredglossary{ignored}
\setabbreviationstyle[ignored]{long-noshort}
\setabbreviationstyle[acronym]{long-short}
\newacronym{api}{API}{application programming interface}
\newacronym{ast}{AST}{abstract syntax tree}
\newacronym{ir}{IR}{intermediate representation}
\newacronym{iv}{IV}{internal variable}
\newacronym{hpc}{HPC}{high performance computing}
\newacronym{dsl}{DSL}{domain-specific language}
\newacronym{gpl}{GPL}{general-purpose language}
\newacronym{glups}{GLUPS}{giga lattice updates per second}
\newacronym{cfd}{CFD}{computational fluid dynamics}
\newacronym{lbm}{LBM}{Lattice Boltzmann method}
\newacronym{fdm}{FDM}{finite difference method}
\newacronym{fvm}{FVM}{finite volume method}
\newacronym{fem}{FEM}{finite element method}
\newacronym{soa}{SoA}{struct of arrays}
\newacronym{aos}{AoS}{array of structure}
\newacronym{pde}{PDE}{partial differential equation}
\newacronym{cse}{CSE}{common subexpression elimination}

\glsdisablehyper

\GlsXtrEnableEntryCounting{acronym}{1}

\makeglossaries

\usepackage{listings}
\usepackage{color}

\ifdefined\PRINTVERSION
 \definecolor{keywordColor}{RGB}{0,0,0}
 \definecolor{commentColor}{RGB}{111,111,111}
 \definecolor{stringColor}{RGB}{69,69,69}
\else
 \definecolor{keywordColor}{RGB}{117,112,179}
 \definecolor{commentColor}{RGB}{27,158,119}
 \definecolor{stringColor}{RGB}{217,95,2}
\fi

\newcommand{\lineNumberStyle}{\tiny\color{gray}}

\ifdefined\FAUPRESS
 \newcommand{\deflstxmargin}{0em} 
\else
 \newcommand{\deflstxmargin}{2.2em}
\fi

\lstdefinelanguage{ExaSlang1}
{
  basicstyle=\ttfamily\small,
  keywords={[2]
    ApplicationHint, ApplicationHints, Cell, DiscretizationHint, DiscretizationHints, Discretize, Domain, Equation, Face\_x, Face\_y, Face\_z, Field, Knowledge, L2Hint, L2Hints, L3Hint, L3Hints, L4Hint, L4Hints, Node, Operator, Solve, SolverHint, SolverHints, \_, all, and, but, cell, coarser, coarsest, current, direction, face\_x, face\_y, face\_z, false, finer, finest, for, generate, import, in, node, not, on, order, solver, to, true, with, %
    waLBerla
  },
  keywordstyle={[2]\color{keywordColor}}, 
  commentstyle=\color{commentColor}, 
  stringstyle=\color{stringColor}, 
  showstringspaces=false,
  sensitive=true, 
  morecomment=[l]{//}, 
  morecomment=[s]{/*}{*/}, 
  morestring=[b]{'}, 
  morestring=[b]{"}, 
  literate={
    {\@}{@}1
  	{\\in}{$\in$}1
  	{\\partial}{$\partial$}1
  	{\\times}{$\times$}1
  	{\\Delta}{$\Delta$}1
  	{\\Omega}{$\Omega$}1
  	{\\pi}{$\pi$}1
  	{\\BSin}{\textbackslash{}in}1
    {\\BSpartial}{\textbackslash{}partial}1
    {\\BStimes}{\textbackslash{}times}1
    {\\BSDelta}{\textbackslash{}Delta}1
    {\\BSOmega}{\textbackslash{}Omega}1
    {\\BSpi}{\textbackslash{}pi}1
  },
  numbers=left,
  numberstyle=\lineNumberStyle,
  xleftmargin=\deflstxmargin,
  captionpos=b,
  escapechar={`},
}

\lstdefinelanguage{ExaSlang2}
{
  basicstyle=\ttfamily\small,
  keywords={[2]
    ApplicationHint, ApplicationHints, Array, Bool, Boolean, Cell, center, Complex, Domain, Double, Equation, Expr, Expression, Face\_x, Face\_y, Face\_z, Field, Float, Globals, Int, Integer, Knowledge, L3Hint, L3Hints, L4Hint, L4Hints, Neumann, Node, None, Operator, Real, Solve, SolverHint, SolverHints, Stencil, StencilTemplate, String, Unit, Val, Value, Var, Variable, all, and, array, bool, boolean, bottom, boundary, but, cell, coarser, coarsest, complex, current, default, double, east, equation, face\_x, face\_y, face\_z, false, finer, finest, float, for, from, generate, i0, i1, i2, import, in, int, integer, is, node, north, not, of, on, operators, prolongation, real, restriction, solver, south, store, string, times, to, top, true, unit, west, with, x, y, z, %
    waLBerla
  },
  keywordstyle={[2]\color{keywordColor}}, 
  commentstyle=\color{commentColor}, 
  stringstyle=\color{stringColor}, 
  showstringspaces=false,
  sensitive=true, 
  morecomment=[l]{//}, 
  morecomment=[s]{/*}{*/}, 
  morestring=[b]{'}, 
  morestring=[b]{"}, 
  literate={
    {\@}{@}1
  },
  numbers=left,
  numberstyle=\lineNumberStyle,
  xleftmargin=\deflstxmargin,
  captionpos=b,
  escapechar={`},
}

\lstdefinelanguage{ExaSlang3}
{
  basicstyle=\ttfamily\small,
  keywords={[2]
    ApplicationHint, ApplicationHints, Array, Bool, Boolean, Cell, center, Complex, Domain, Double, Equation, Expr, Expression, Face\_x, Face\_y, Face\_z, Field, Float, Func, FuncTemplate, Function, FunctionTemplate, Globals, Inst, Instantiate, Int, Integer, Knowledge, L2, L4Hint, L4Hints, Neumann, Node, None, Operator, Real, Stencil, String, Unit, Val, Value, Var, Variable, all, and, append, array, as, bc, bool, boolean, bottom, boundary, but, cell, coarser, coarsest, color, complex, count, current, default, double, east, else, face\_x, face\_y, face\_z, false, finer, finest, float, for, from, generate, i0, i1, i2, if, import, in, int, integer, jacobi, locally, loopBase, modifiers, node, north, not, of, on, or, override, prepend, prolongation, real, relax, repeat, replace, restriction, return, smootherHint, smootherStage, solve, solveFor, solver, south, string, times, to, top, true, unit, until, west, where, while, with, x, y, z, %
    waLBerla
  },
  keywordstyle={[2]\color{keywordColor}}, 
  commentstyle=\color{commentColor}, 
  stringstyle=\color{stringColor}, 
  showstringspaces=false,
  sensitive=true, 
  morecomment=[l]{//}, 
  morecomment=[s]{/*}{*/}, 
  morestring=[b]{'}, 
  morestring=[b]{"}, 
  literate={
    {\@}{@}1
  },
  numbers=left,
  numberstyle=\lineNumberStyle,
  xleftmargin=\deflstxmargin,
  captionpos=b,
  escapechar={`},
}

\lstdefinelanguage{ExaSlang4}
{
  basicstyle=\ttfamily\small,
  keywords={[2]
    Array, Bool, Boolean, CVector, Cell, center, ColumnVector, Complex, Domain, Edge\_Cell, Edge\_Node, Equation, Expr, Expression, Face\_x, Face\_y, Face\_z, Field, Func, FuncTemplate, Function, FunctionTemplate, Globals, Inst, Instantiate, Int, Integer, Knowledge, Layout, LayoutTransformations, Matrix, Neumann, Node, None, Operator, RVector, Real, RowVector, Set, Stencil, StencilField, StencilTemplate, String, T, Unit, Val, Value, Var, Variable, Vec2, Vec3, Vec4, Vector, active, activeSlot, advance, all, and, apply, as, bc, begin, bottom, boundary, break, but, cell, coarser, coarsest, color, communicate, communicating, communication, concat, contraction, count, current, currentSlot, default, dup, east, edge\_cell, edge\_node, else, ending, external, face\_x, face\_y, face\_z, false, finer, finest, finish, fragments, from, fromFile, ghost, i0, i1, i2, if, import, into, jacobi, locally, loop, next, nextSlot, node, noinline, north, not, of, on, only, or, over, postcomm, precomm, previous, previousSlot, prolongation, reduction, relax, rename, repeat, restriction, return, sequentially, solve, south, starting, stepping, steps, times, to, top, transform, true, until, west, where, while, with, x, y, z, %
    waLBerla
  },
  keywordstyle={[2]\color{keywordColor}}, 
  commentstyle=\color{commentColor}, 
  stringstyle=\color{stringColor}, 
  showstringspaces=false,
  sensitive=true, 
  morecomment=[l]{//}, 
  morecomment=[s]{/*}{*/}, 
  morestring=[b]{'}, 
  morestring=[b]{"}, 
  literate={
    {\@}{@}1
  },
  numbers=left,
  numberstyle=\lineNumberStyle,
  xleftmargin=\deflstxmargin,
  captionpos=b,
  escapechar={`},
}

\lstdefinelanguage{Config}
{
  basicstyle=\ttfamily\small,
  keywords={[2]
    import, true, false
  },
  keywordstyle={[2]\color{keywordColor}}, 
  commentstyle=\color{commentColor}, 
  stringstyle=\color{stringColor}, 
  showstringspaces=false,
  sensitive=true, 
  morecomment=[l]{//}, 
  morecomment=[s]{/*}{*/}, 
  morestring=[b]{'}, 
  morestring=[b]{"}, 
  numbers=left,
  numberstyle=\lineNumberStyle,
  xleftmargin=\deflstxmargin,
  captionpos=b,
  escapechar={`},
}

\lstdefinestyle{MyScala}{
  language=scala,
  basicstyle=\ttfamily\small,
  showstringspaces=false,
  keywordstyle=\color{keywordColor}, 
  commentstyle=\color{commentColor}, 
  stringstyle=\color{stringColor}, 
  numbers=left,
  numberstyle=\lineNumberStyle,
  xleftmargin=\deflstxmargin,
  captionpos=b,
  escapechar={`},
}

\lstdefinelanguage{MyCpp}{
  language=C++,
  basicstyle=\ttfamily\small,
  showstringspaces=false,
  keywordstyle=\color{keywordColor}, 
  commentstyle=\color{commentColor}, 
  stringstyle=\color{stringColor}, 
  numbers=left,
  numberstyle=\lineNumberStyle,
  xleftmargin=\deflstxmargin,
  captionpos=b,
  escapechar={`},
}

\lstdefinelanguage{MyCMake}{
  basicstyle=\ttfamily\small,
  showstringspaces=false,
  keywordstyle=\color{keywordColor}, 
  commentstyle=\color{commentColor}, 
  stringstyle=\color{stringColor}, 
  numbers=left,
  numberstyle=\lineNumberStyle,
  xleftmargin=\deflstxmargin,
  captionpos=b,
  escapechar={`},
}

\lstdefinelanguage{MyCMakeCoupling}{
  language=MyCMake,
  keywords={[2]
  waLBerla_generate_target_from_exaslang, NAME, PATH, PLATFORM, KNOWLEDGE
  },
}

\makeatletter
\def\addToLiterate#1{\edef\lst@literate{\unexpanded\expandafter{\lst@literate}\unexpanded{#1}}}
\lst@Key{add to literate}{}{\addToLiterate{#1}}
\makeatother

,commentstyle={},stringstyle={},add to literate={\\-}{}{0\discretionary{-}{}{}}]}
,commentstyle={},stringstyle={},add to literate={\\-}{}{0\discretionary{-}{}{}}]}
,commentstyle={},stringstyle={},add to literate={\\-}{}{0\discretionary{-}{}{}}]}
,commentstyle={},stringstyle={},add to literate={\\-}{}{0\discretionary{-}{}{}}]}

,commentstyle={},stringstyle={},add to literate={\\-}{}{0\discretionary{-}{}{}}]}

\lstnewenvironment{code}[1][]%
{
   \lstset{float,belowskip=-\baselineskip,#1}}
{
}

\usepackage[nameinlink, capitalise]{cleveref}

\crefname{equation}{Eq.}{Eq.}
\Crefname{equation}{Equation}{Equations}
\crefname{figure}{Fig.}{Fig.}
\Crefname{figure}{Figure}{Figures}
\crefname{table}{Tab.}{Tab.}
\Crefname{table}{Table}{Tables}
\crefname{lstlisting}{Lst.}{Lst.}
\Crefname{lstlisting}{Listing}{Listings}
\crefname{section}{Sect.}{Sect.}
\Crefname{section}{Section}{Sections}

\crefformat{footnote}{#2\footnotemark[#1]#3}

\usepackage[export]{adjustbox}

\usepackage{floatrow}

\usepackage[justification=centering]{caption}
\usepackage{subcaption}

\usepackage{tikz}
\newcommand\TikCircle[1][1]{\tikz[baseline=-#1]{\draw[very thick](0,0.075)circle[radius=#1mm];}}

\usepackage{ctable} 
\usepackage{makecell} 
\usepackage{multirow}

\usepackage{amssymb}
\usepackage{marvosym}

\usepackage[shortcuts]{extdash}
\begin{document}

\author[a, $\ast$]{Richard Angersbach\orcidlink{0000-0001-7749-8239}}
\author[a,b]{Sebastian Kuckuk\orcidlink{0000-0002-6782-3162}}
\author[a,b]{Harald K\"ostler\orcidlink{0000-0002-6992-2690}}
\affil[a]{Chair for Computer Science 10 (System Simulation), Friedrich-Alexander-Universität Erlangen-Nürnberg, Cauerstraße 11, 91058 Erlangen, Germany}
\affil[b]{Erlangen National High Performance Computing Center (NHR@FAU), Martensstraße 1, 91058 Erlangen, Germany}
\affil[$\ast$]{Contact R. Angersbach. Email: \texttt{richard.angersbach@fau.de}}

\title{Towards Code Generation for Octree-Based Multigrid Solvers}

\maketitle

\begin{abstract}
\par
This paper presents a novel method designed to generate multigrid solvers optimized for octree-based software frameworks.
Our approach focuses on accurately capturing local features within a domain while leveraging the efficiency inherent in multigrid techniques.
We outline the essential steps involved in generating specialized kernels for local refinement and communication routines, integrating on-the-fly interpolations to seamlessly transfer information between refinement levels.
For this purpose, we established a software coupling via an automatic fusion of generated multigrid solvers and communication kernels with manual implementations of complex octree data structures and algorithms often found in established software frameworks.
We demonstrate the effectiveness of our method through numerical experiments with different interpolation orders.
Large-scale benchmarks conducted on the SuperMUC-NG CPU cluster underscore the advantages of our approach, offering a comparison against a reference implementation to highlight the benefits of our method and code generation in general.
\end{abstract}
\section{Introduction}\label{sec:intro}
\par
\Glspl{pde} are ubiquitous in modeling physical phenomena from various scientific application domains.
The multigrid methods offer an efficient solution for solving the resulting systems of equations with complexity $O(N)$, where N is the number of unknowns.
Solving these equations often requires high resolutions, especially near complex geometries, but employing fine resolutions throughout the whole computational domain can be wasteful or infeasible.
Using k-d trees for space partitioning allows for local refinement capabilities.
Leveraging multigrid methods on refined meshes combines adaptability with efficiency.
Our work aims to explore the use of code generation technology for generating multigrid solvers on refined block-structured meshes,
with a focus on improving communication regarding interpolation order and performance and enabling the generation of kernels specialized to octree-based refinement.
\par
This work is structured as follows.
Our employed simulation software stack is described in~\cref{sec:software}.
In~\cref{sec:refined_comm}, we showcase our communication concept for meshes with refinement.
We also describe the necessary adaptations to our code generation pipeline in~\cref{sec:codegen}.
The capabilities of our approach are shown for a locally refined 3D Poisson problem in~\cref{sec:app}.
\Cref{sec:bench} contains the results of scaling experiments conducted on the SuperMUC-NG CPU cluster.
\section{Software}\label{sec:software}

To generate multigrid solvers on domains with mesh refinement, data structures like octrees had to be made accessible to the code generation pipeline.
These data structures are usually highly complex and are often implemented in software frameworks written in~\glspl{gpl} such as C/C++.
We bridge the gap by acquainting the code generator with mesh refinement concepts from other frameworks, making them accessible for the generated solvers via automated interfaces.
This approach simplifies the generation process by utilizing existing implementations for complex data structures.
This work builds upon an existing coupling strategy between the waLBerla and ExaStencils frameworks~\cite{exa_wb_coupling}.
\par
ExaStencils~\cite{exa_p2} offers the generation of whole programs in C++ for geometric multigrid solvers from its own external \gls{dsl} called ExaSlang.
ExaSlang follows a multi-layered language approach, with each of the four layers tailored for domain experts from different communities~\cite{exaslang}.
ExaStencils employs a Scala-based source-to-source compiler to parse ExaSlang sources, apply code transformations for automatic parallelization and optimization, and pretty-print C++ code.
The generated code supports automatic parallelization with OpenMP, MPI, and CUDA, along with optimizations such as \gls{cse}, address pre-calculation, SIMD vectorization, and polyhedral loop transformations.
\par
waLBerla~\cite{walberla} is a modular C++17 framework dedicated to \gls{cfd}, especially for the \gls{lbm}.
Code generation of single kernels is enabled via the Python modules pystencils and lbmpy.
pystencils, leveraging SymPy, generates optimized stencil codes with automatic SIMD vectorization, \gls{cse}, and non-temporal stores.
lbmpy extends pystencils with further abstractions for the generation of efficient \gls{lbm} kernels.
waLBerla supports MPI for distributed-memory parallelism and OpenMP for shared-memory parallelism.
The execution on AMD and NVIDIA GPU platforms is facilitated via HIP and CUDA~\cite{thermocapillary}.
\par
While both frameworks generally employ block-structured domain partitioning, their properties and implementation details differ significantly.
\begin{figure}
    \centering
    \includegraphics[width=0.6\linewidth]{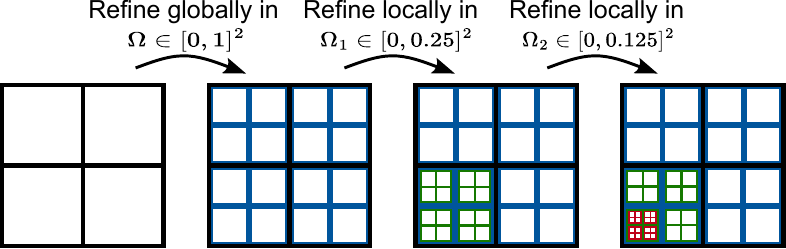}
    \caption{Exemplary 2D mesh refinement in waLBerla. From left to right the grid is refined in the lower left corner.}
    \label{fig:partitioning}
\end{figure}
\par
ExaStencils employs a hierarchical domain partitioning approach inspired by parallelism across hardware layers and supports a spectrum of grids ranging from uniform to non-uniform grids~\cite{exa_automatic, Kuckuk2019}.
Domains are organized into blocks and fragments.
Blocks, the top hierarchy level, are mapped to distributed-memory parallel entities like MPI processes.
Each block can be divided into equally sized fragments, each exclusively linked to a block.
Fragments represent parts of the computational grid and are often mapped to sockets, cores, or accelerators, facilitating another layer of parallelization.
\par
In waLBerla, an octree-based domain partitioning is used~\cite{walberla}.
Initially, a cuboidal domain is regularly divided into equally-sized sub-domains, each serving as the root of an octree.
These root blocks collectively form the blockforest, waLBerla's central data structure, and can undergo further recursive refinement.
\Cref{fig:partitioning} illustrates this procedure on a two-dimensional unit domain $\Omega$ with four root blocks being incrementally refined.
The refinement levels are depicted by colored contours, with red representing level three, green level two, and blue level one.
Starting from a refinement level of zero on the left-most side, all root blocks are first globally refined once, followed by a local refinement in $\Omega_{1}$ and $\Omega_{2}$, respectively.
The refinement levels are depicted by colored contours, with red representing level three, green level two, and blue level one.
These refinement selection steps are commonly specified in user-defined functions or in custom configuration files.
Once the refinement is completed, the blocks at the leaves of this structure represent the final sub-domains.
In the context of distributed memory parallelism, each block is assigned exclusively to one process, although a process may manage multiple blocks.
The only restriction for the refinement is a 2:1 size ratio between adjacent blocks.
\par
The software coupling between both frameworks was limited to regular domain partitioning of rectangular domains into uniform, structured building blocks.
This required an exact match between the partitioning used in both framework codes, as ExaStencils lacked familiarity with mesh refinement.
Our aim in this work is to surpass this limitation and explore the necessary steps to extend our code generator to operate on refinement data structures from waLBerla.

\section{Refined Communication}\label{sec:refined_comm}
\par
This chapter presents our approach to exchanging data between blocks of varying refinement levels.
As a demonstration, we use a 2D non-periodic domain that is initially divided into two root blocks \texttt{A} and \texttt{B},
root block \texttt{A} undergoes further refinement into four child blocks \texttt{A0} to \texttt{A3}.
Each block has the same amount of cells, leading to a non-uniform discretization of the domain.
\subsection{Data Exchange}
\par
\begin{figure}
    \centering
    \includegraphics[width=1.0\linewidth]{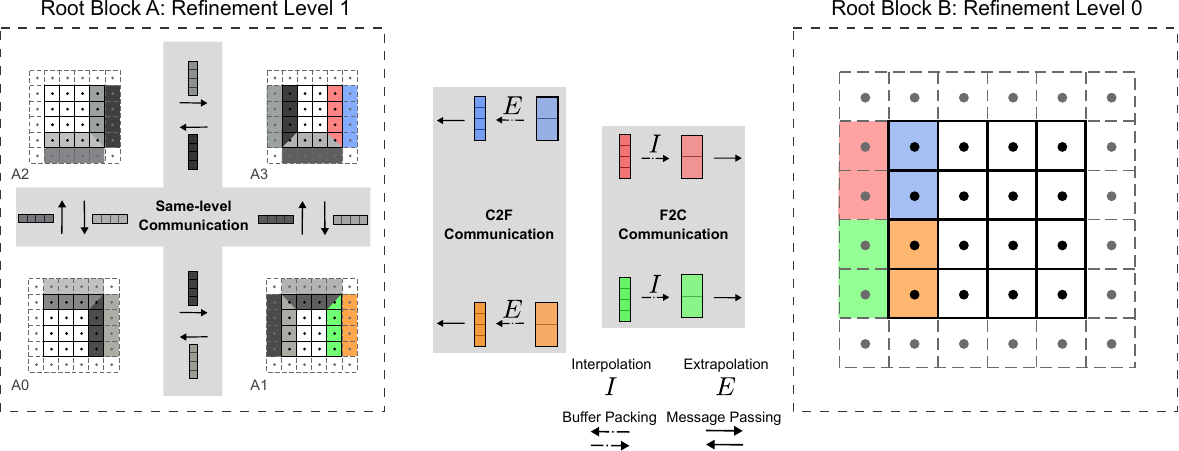}
    \caption{Data exchange for communication for our exemplary 2D domain setup.}
    \label{fig:communication}
\end{figure}

\Cref{fig:communication} illustrates the data flow of our communication scheme,
categorized into three refinement cases:
same-level communication among all child blocks (colored in grey shades),
coarse-to-fine (C2F) exchange from \texttt{B} to \texttt{A1} and \texttt{A3} (orange/blue),
and fine-to-coarse (F2C) communication from \texttt{A1} and \texttt{A3} to \texttt{B}~(green/red).
\par
In this work, we will put high emphasis on the C2F and F2C cases.
The same-level ghost layer exchange from~\cite{exa_automatic} remains unchanged.
From~\cref{fig:communication}, it becomes evident that coinciding cell regions between the coarse block and its adjacent fine blocks must be determined for both cases.
For a block with level $L_{curr}$ and neighbor in one of the cardinal directions with level $L_{neigh}$,
a refinement ratio $r$ and dimensionality $D$, we define the number of potential neighbors as
\begin{equation*}
    N_{neigh} = \begin{cases} r^{D - 1} & \text{if } L_{curr} < L_{neigh} \\ 1 & \text{otherwise} \end{cases}.
\end{equation*}
For our 2:1 refinement balance, a coarse block has two fine neighbor blocks in 2D and four in 3D.
For C2F communication, we split the iteration space of the coarse block's inner cells at the block interface (edge in 2D, face in 3D) into $N_{neigh}$ segments, each for an adjacent fine block.
The splitting divides each axis of the block interface into $r$ equally-sized pieces.
With our given $r = 2$ ratio, a coarse edge (2D) with 4 cells splits into two segments with 2 cells each, while a coarse face (3D) with $4 \times 4$ cells splits into four segments with $2 \times 2$ cells each.
Each coarse segment's inner cell values must then be communicated to the unsplit ghost layers of a fine block,
e.g. the inner values from the two blue cells of block \texttt{B} must be mapped to the ghost values of the four blue cells in \texttt{A3}.
Similarly, the inner cells of an unsplit fine block must be synchronized with the corresponding segment of ghost layers of a coarse block,
e.g. the inner values from the four green cells of block \texttt{A1} must be mapped to the ghost values of the two green cells in \texttt{B}.
For C2F mappings, we extrapolate coarse block data, while for F2C mappings, we interpolate fine block data.
\subsection{Message Passing}
\par
\Cref{fig:communication} also shows the differences in message passing for the refinement cases.
Same-level communication does not necessarily require communication buffers for (un-)packing of data.
Instead, MPI concepts like derived datatypes can bypass \emph{explicit} copy operations and directly send values from one memory location to another.
However, in the other refinement cases, not only copies but also mappings with inter-/extrapolation may be needed.
In this case, we directly store the mapping results in the communication buffers to be sent and unpacked into the ghost layers of the neighbor blocks.
Compared to same-level communication, we potentially have $N_{neigh}$ neighbor blocks per cardinal direction.
Establishing an association between a message and its corresponding segment is crucial for distinguishing received messages.
For instance, block \texttt{B} receives two messages from adjacent fine blocks, storing them in receive buffers.
This association is then also used to unpack the values from the receiving buffers into the correct ghost layer segments.
In~\cref{fig:communication}, block \texttt{B} stores two fine send buffers (shaded in orange/blue) and two coarse receive buffers (red/green) for the west direction, mapped to corresponding cell segments.
\texttt{A1} and \texttt{A3}, however, only need one coarse send buffer and one fine receive buffer.
To coordinate the message transmission to correct communication buffers, we encode the segment index of the neighbor block into the MPI tag of each message.

\subsection{Interpolation/Extrapolation}
For macroscopic fields, waLBerla initially offered a manual implementation, mandating two ghost layers at each dimension's start and end.
The interpolations employed for C2F and F2C communication had a constant and linear order, respectively.
This implementation not only risks accuracy degradation due to low-order interpolations but also wastes memory, especially with smaller block sizes (e.g., 4 cells per dimension) common in mesh refinement and multigrid.
In ExaStencils, we redesigned communication to support various interpolation schemes for the available variable localizations.
Focusing on waLBerla's cell-centered variables, we propose an approach with quadratic C2F and linear F2C schemes, requiring only half of the ghost layers per dimension.
\par
\begin{figure}
    \begin{adjustbox}{minipage=\linewidth, scale=1.1}
        \hspace*{-0.05\linewidth}
        \begin{subfigure}[b]{0.28\linewidth}%
            \centering
            \includegraphics[width=1.00\linewidth]{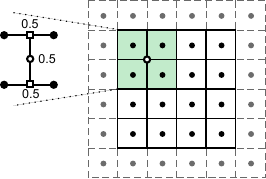}
            \caption{Linear F2C scheme.}
            \label{subfig:f2c_2d}
        \end{subfigure}
        \hfill%
        \begin{subfigure}[b]{0.68\linewidth}%
            \centering
            \includegraphics[width=1.00\linewidth]{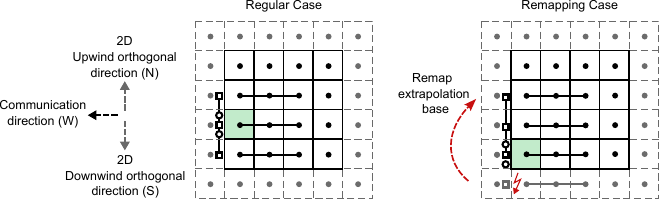}
            \caption{Quadratic C2F scheme.}
            \label{subfig:c2f_2d}
        \end{subfigure}
    \end{adjustbox}
    \caption{Proposed extra-/interpolation schemes for different refinement cases.}
    \label{fig:schemes}
\end{figure}

While it is possible to formulate interpolation schemes requiring data from adjacent fine and coarse blocks,
this work focuses on a highly parallelizable approach using only the inner cell data stored in the sending block, i.e. completely agnostic from a neighbor's values.
\Cref{fig:schemes} depicts the inner cells used as support for our proposed linear F2C and quadratic C2F interpolation schemes.
\subsubsection{F2C Interpolation}
\par
The linear interpolation scheme for F2C communication in the west direction $d_{comm}$ is depicted in~\cref{subfig:f2c_2d}.
Fine cells overlapping with a coarse ghost cell (shaded in green) are determined for computing the coarse value using bilinear (2D) or trilinear (3D) interpolation.
For uniform blocks, the interpolation weights for fine values are known constants, e.g. $0.25$ in 2D and $0.125$ in 3D.
In the F2C packing routine, we use a stepsize of $r$ in each dimension of the block interface for iterating over fine inner cells,
compute the coarse cell values and store them in the communication buffer.
The adjacent coarse block can then unpack the received buffer directly into its ghost layer segment.
\subsubsection{C2F Extrapolation}
\par
\begin{figure}
    \centering
    \includegraphics[width=0.75\linewidth]{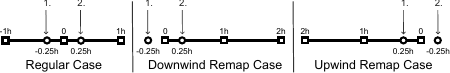}
    \caption{Weight computation for three different remapping cases.}
    \label{fig:c2f_cases}
\end{figure}

Besides the F2C approach, we now have fewer cells in a coarse block segment than ghost layer cells in a fine block and address this data deficit with extrapolation.
While the order of our C2F extrapolation method can be adapted by changing the number of neighbor cells used as bases,
this chapter concentrates on a quadratic C2F scheme employing 3 coarse cells as bases.
\paragraph*{Two-dimensional scheme:}
A schematic view of our quadratic C2F mapping in 2D is shown in~\cref{subfig:c2f_2d} with a communication in the west direction $\vec{d}_{comm}$.
We employ second-degree Lagrange polynomials to compute all bases that are depicted as hollowed shapes.
At first, we compute a base (depicted as a hollowed square \Squarepipe) for our current coarse cell (shaded in green) by performing a quadratic extrapolation.
The bases for this extrapolation comprise values from the current coarse cell and the two coarse neighbor cells in the inverse communication direction $-\vec{d}_{comm}$ (here: east).
With our approach, we expect the blocks to have at least three cells per dimension.
We will denote this step as the extrapolation in the first dimension (here: the x-dimension).
To go beyond a one-dimensional extrapolation, we repeat this step for adjacent cells to compute further extrapolated bases (\Squarepipe) that are then used for an extra-/interpolation in the second dimension (here: the y-dimension).
We determine the location of these bases with the set of the cardinal directions $C_{2D} = \{W, E, S, N\}$ where $W$ stands for the (downwind) west direction, $E$ for the (upwind) east direction, etc.
From those, we exclude the communication direction and its inverse to obtain a set of orthogonal directions $O_{2D} = C_{2D} \, \setminus \, \{\vec{d}_{comm}, -\vec{d}_{comm}\}$.
The orthogonal directions in this example would be $O_{2D} = \{S, N\}$.
Here, we distinguish between 2D orthogonal vectors in upwind $\vec{o}_{2D}^{\, +} = N$ and downwind direction $\vec{o}_{2D}^{\, -} = S$.
The location of the additional extrapolated bases (\Squarepipe) is then determined by offsetting the current cell with the directions from $O_{2D}$ individually.
However, one must be cautious to avoid building bases in ghost layer regions with outdated data.
On the left-hand side of~\cref{subfig:c2f_2d}, all orthogonal neighbor cells are within the inner cell regions of the block.
On the right-hand side, however, the orthogonal neighbor in the downwind direction would build a base within the ghost layer regions.
We circumvent this issue by introducing a remapping mechanism that checks if a neighbor in orthogonal direction $\vec{o}$ is in the ghost region and, in this case, applies a remapping to an inner cell in direction $\vec{o}_{remap} = -2 \vec{o}$.
We now span an axis across all extrapolated bases (\Squarepipe) in the direction of the 2D orthogonal vectors to perform an extra-/interpolation in the second dimension.
Note that for communication in the south direction, the extrapolations in the first and second dimensions would be in y-dimension and x-dimension, respectively.
The Lagrange weights can be computed as
\[
w_i(x) = \prod_{j=0, j\neq i}^{n-1} \frac{x - \mathbf{x}_j}{\mathbf{x}_i - \mathbf{x}_j}
\]
where $x$ is our target location on the axis and $\mathbf{x} = \{x_0, \ldots, x_{n-1}\}$ contains the locations of the $n = 3$ bases.
The target value of the fine cell can then be evaluated as the weighted sum between the values of the bases and the computed weights.
It becomes apparent that this remapping also has implications for the weight computation as the position of the bases changes.
Here, we distinguish between three different remapping cases as depicted in~\cref{fig:c2f_cases}.
\par
\begin{figure}
    \centering
    \includegraphics[width=1.0\linewidth]{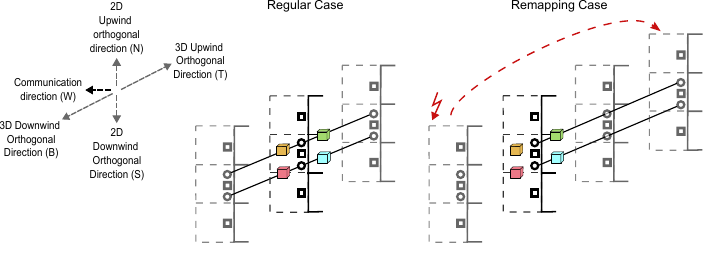}
    \caption{3D quadratic C2F extra-/interpolation scheme.}
    \label{fig:c2f_3d}
\end{figure}

The first case on the leftmost side is without remapping.
Here, we can compute the two fine cell values (depicted as hollowed circles $\TikCircle$) by performing an interpolation on the axis with the bases $\mathbf{x} = \{-h, 0, h\}$ with $h$ being the cell width.
The locations of the fine cell values are $-0.25h$ for the first (downwind) and $0.25h$ for the second (upwind).
The second case is illustrated in the middle and shows the position of the bases and the fine cell values when a remapping of the downwind orthogonal direction occurs.
Relative to our current coarse cell, we now have the bases $\mathbf{x} = \{0, h, 2h\}$ where the remapped base resides at $2h$.
The location of our first fine cell value at $-0.25h$ is outside of the span of the bases which again implies that an extrapolation is done.
The second fine cell value at $0.25h$ is on the axis again and can be computed via interpolation.
The third case on the right-hand side is essentially a mirroring of the second case where the upwind orthogonal direction is remapped and, correspondingly, the first fine value is interpolated and the second is extrapolated.
\paragraph*{Three-dimensional scheme:}
The C2F implementation for 3D scenarios builds on top of the 2D implementation, i.e. the steps for the extra-/interpolations in the first and second dimensions are identical.
In 3D, we now have the set of cardinal directions $C_{3D} = \{W, E, S, N, B, T\}$ which also extends the set of orthogonal directions $O_{3D}$ to have four entries.
For an exemplary communication direction $W$, we have the set $O_{3D} = \{S, N, B, T\}$ where the 3D upwind and downwind orthogonal directions are $\vec{o}_{3D}^{\, +} = T$ and $\vec{o}_{3D}^{\, -} = B$, respectively.
As shown in~\cref{fig:c2f_3d}, extra-/interpolation in the third dimension will consider the 2D bases ($\TikCircle$) from three 2D slices of the field,
namely from the current cell and the neighbors in both 3D orthogonal directions.
To compute the 2D bases for these neighbors, we first have to compute the 1D extrapolated bases (\Squarepipe) for the slice in direction $\vec{o}_{3D}^{\, +}$ at locations computed by building diagonals with the 2D orthogonal vectors, i.e. at $\{\vec{o}_{3D}^{\, +} - \vec{o}_{2D}^{\, +}, \vec{o}_{3D}^{\, +}, \vec{o}_{3D}^{\, +} + \vec{o}_{2D}^{\, +} \}$.
The same procedure is applied for the slice in direction $\vec{o}_{3D}^{\, -}$.
Note that remapping of the location of a slice can occur to avoid building bases with ghost layer values.
With the 1D bases, we can finally build the 2D bases ($\TikCircle$) in the neighboring slices where each slice holds an upwind and a downwind value for the 2D orthogonal directions.
In total, we now have six 2D bases ($\TikCircle$) which we will use to compute four values for the adjacent fine cell.
We do so by spanning an axis over the downwind values of the slices and another axis over the upwind values.
For each axis, we extra- or interpolate again depending on the three cases in~\cref{fig:c2f_cases}.
Combining the cases of the 2D and 3D extra-/interpolations, we possibly have nine different computation rules for our quadratic scheme,
namely one for each corner of an interface with remappings in two dimensions,
one for each edge of an interface with remapping in one dimension,
and a regular case without remapping for the remaining cells.
Since the iteration spaces of the coarse blocks are split into segments, some of the cases become unattainable.
With introspection of the iteration spaces, our code generator eliminates the dead code for these cases.
\par
\begin{figure}
    \centering
    \includegraphics[width=0.65\linewidth]{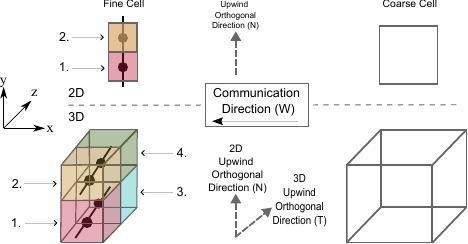}
    \caption{Packing ordering example for 2D and 3D.}
    \label{fig:c2f_order}
\end{figure}

The computed $N_{neigh}$ fine values per coarse cell are then packed into the communication buffer as shown in~\cref{fig:c2f_order}.
In 2D, the values are packed in the same order as they are computed in~\cref{fig:c2f_cases}.
In 3D, we have two axes that are offset in the 2D upwind orthogonal direction.
Each axis spans over the neighbors in the direction of the 3D orthogonal vectors and computes a downwind and upwind value.
We first pack the downwind value of the axis below, followed by the downwind value of the upper axis and repeat that for the upperwind values.
Note that the adjacent fine block has the same set of orthogonal directions $O$ which allows us to reuse this protocol when unpacking the buffer into the ghost layer region.
When iterating over the edge or face of fine ghost cells, we employ a step size of $r$ in each dimension.
In 2D, we then first unpack the value into the current cell of the iteration and the second is stored in the cell offset by $\vec{o}_{2D}^{\, +}$.
This also applies to the first two values in 3D.
The third value is stored at the offset $\vec{o}_{3D}^{\, +}$ and the fourth uses an offset in the diagonal vector between the upwind directions $\vec{o}_{2D}^{\, +} + \vec{o}_{3D}^{\, +}$.
\paragraph*{Communication volume:}
\par
One important goal of communication is to minimize the volume of data that is communicated to the neighbor blocks.
There exist variants where the values are first interpolated before sending messages and variants where interpolation occurs after receiving messages.
The most optimal variant strongly depends on the interpolation method and its order since this determines the number of coarse cells required as support.
A constant-order scheme benefits from sending coarse values first which are then directly propagated to the fine values.
Our second-order scheme requires the support of three coarse values which would benefit from performing the communication beforehand.
In this work, we established a modular software design that allows us to choose the best variant for communication volume efficiency.
\section{Code Generation}\label{sec:codegen}

This chapter covers the necessary steps for familiarizing a code generator with mesh refinement concepts from another framework
and shows which adaptations were required to support our proposed communication scheme from~\cref{sec:refined_comm}.

\subsection{Data Structures}
\par
Introducing mesh refinement concepts to an existing codebase causes major changes to the core data structures.
Without mesh refinement, the number of potential neighbor blocks per dimension is known beforehand which allows for a static allocation of data structures for geometric and topologic information.
In ExaStencils, this information is stored in so-called \glspl{iv}~\cite{Kuckuk2019}.
One example for them is a boolean variable storing whether an adjacent fragment is on a remote MPI process and is used to distinguish between remote and local communication schemes.
To support refined meshes with a variable number of blocks per process and a variable number of neighbors, we now rely on dynamic data structures such as C++ standard vectors.
Topological information is now stored in an additional array dimension for the $N_{neigh}$ neighbor blocks.
Within our generator, we introduced wrapper data structures with dynamic data allocation that are automatically employed in the case of mesh refinement.
In our software coupling, these data structures are set up when the interface between the generated solver and the framework code is initialized.

\subsection{Operators}
\par
Operators for numerical solvers usually require geometric information such as the cell width of a block as shown in the operator declaration for a gradient approximation in~\cref{lstlisting:grad_decl}.
Depending on the properties of a block, this information can either be expressed with a simple expression (or even a constant value) in the case of uniform blocks or must be stored explicitly in the case of non-uniform blocks.
For this purpose, ExaStencils provides abstractions in the \gls{dsl}, the so-called virtual fields~\cite{Kuckuk2017towards}, that obfuscate storage details and let the generator choose the most efficient case to obtain this information.
Despite using uniform building blocks, having locally refined resolutions implies having a non-uniform discretization of the domain, which requires the operators to also consider refinement information.
In this work, we want to go beyond the existing concepts and establish an automatic re-discretization of an operator for refinement without changing the user interface.
We achieved that by implementing wrapper data structures for virtual fields that automatically fetch this information via accessors provided by a waLBerla block.

\begin{code}[
    caption={Exemplary operator declaration in ExaSlang 4 for a finite-difference gradient approximation in x-direction},
    label={lstlisting:grad_decl},
    language=ExaSlang4,
    ]
Stencil GradX@all {
  [-1, 0, 0] => -1. / (2. * vf_cellWidth_x),
  [ 1, 0, 0] =>  1. / (2. * vf_cellWidth_x)
}
\end{code}

\subsection{Loops}

With the addition of dynamic data structures, the bounds of the loop over the local blocks of an MPI process become unknown.
To enable specialized kernels for the potentially different resolutions for each block, we have enabled a loop invariant motion mechanism to minimize the time spent fetching common scalar data structures from waLBerla.
We introduced the concept of process-local loop variables that are automatically fetched in a loop over the blocks when present in the \gls{ast} of a kernel.
This concept is also highly beneficial for generating GPU kernels since the promoted scalars can be directly passed to the kernel.
For our exotic use case of employing multigrid with potentially very small block sizes, we have also implemented an optimization strategy for pre-fetching pointers to waLBerla data structures to avoid type-checking overheads employed by regular waLBerla accessors which are usually no concern in the scope of LBM.

\begin{code}[
  caption={Simplified code example for automatic loop invariant motion of refinement information.},
  label={lstlisting:loop_invariant_motion},
  language=MyCpp,
  ]
for (int bIdx = 0; bIdx < localBlocks.size(); ++bIdx) {
  auto block = localBlocks[bIdx];
  size_t refinementLvl = blockforest->getLevel(*block);
  real_t dx = blockforest->dx(refinementLvl);

  // ... kernel code
}
\end{code}

\subsection{Communication}
\par
As a proof of concept, we first developed our communication scheme manually in close collaboration with the waLBerla developers as a baseline.
In this work, we want to investigate the benefits of generating such complex routines and therefore discuss the steps that are necessary for integrating them into our software coupling.
The existing communication implementation in ExaStencils provided classes encapsulating local and remote communication kernels where the latter commonly employ additional kernels for the (un-)packing of values to/from buffers.
All these classes must be extended to support the different refinement cases with potentially varying communication volumes, interpolation orders, and variable localizations within a mesh.
We realize this by introducing a modular design with abstraction layers for each case which is then mapped to efficient implementations by the generator.
As an optimization, we employ automatic communication hiding by overlapping the local with the remote communication to avoid targeting the same main memory bottleneck that would occur when overlapping computation with local communication.
\section{Application}\label{sec:app}

\begin{figure}
    \centering
    \includegraphics[width=0.35\linewidth]{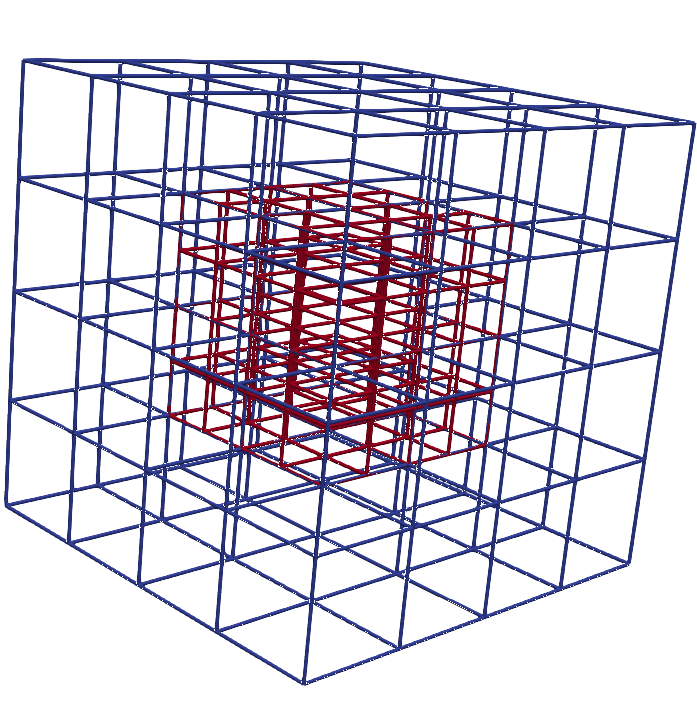}
    \caption{Domain with local refinement in the center.}
    \label{fig:grid_part}
\end{figure}

This chapter investigates the numerical accuracy of multigrid solvers on octrees with different C2F schemes.
We consider a standard benchmark problem given by the 3D Poisson equation~(\cref{eq:1}) with the Laplacian operator $\Delta$, right-hand side function $f(x,y,z) = 0$, and Dirichlet boundary conditions~(\cref{eq:2}).
\begin{align}
    -\Delta u(x,y,z) = f(x,y,z) & \qquad x,y,z \text{ on } \Omega \label{eq:1} \\
    u(x,y,z) = sin(\pi \cdot x) \cdot sin(\pi \cdot y) \cdot sinh(\sqrt{2} \cdot \pi \cdot z) & \qquad x,y,z \text{ on } \partial \Omega \label{eq:2}
\end{align}
We discretize the continuous problem on the computational domain $\Omega = [0,1]^3$ using the cell-centered \gls{fvm} and solve the arising linear system with a $V(3, 3)$ multigrid solver~\cite{Kuckuk2019}.
The solver uses a damped Jacobi smoother with $\omega = 0.8$, serving also as the coarse-grid solver.
We terminate the solver when the $L_2$ residual norm falls below a threshold of $1\mathrm{E}{-16}$, or when the maximum of 35 iterations is reached.
The obtained numerical solution $u^*$ is then used to compute the $L_2$ error norm as $\left\Vert u^* - u \right\Vert_2$.
We compare errors across different resolutions using the block partitioning visualized in~\cref{fig:grid_part}:
it consists of 56 blocks on refinement level 2 (blue) and 64 blocks on level 3 (red).
\par
\Cref{table:err_analysis} summarizes the obtained results.
The given block size is equal to the number of cells per block on the finest multigrid level.
Coarser levels are added until a local resolution of $4^3$ cells is reached.
Apart from the error norm itself, its grid convergence $\kappa$~(\cref{eq:3}) is a vital metric.
We expect quadratic convergence, that is when halving the local cell size $h$, $\kappa$ should be $0.25$, as we employ a second-order discretization of our \gls{pde}.
\begin{equation} \label{eq:3}
    \kappa = \frac{\Vert u^*_{h/2} - u_{h/2} \Vert_2}{\Vert u^*_h - u_h \Vert_2}
\end{equation}
\par
We can see two effects.
One, the absolute error decreased with increasing interpolation orders.
Second, and more importantly, the chosen interpolation order limits the convergence order (constant for constant, linear for first-order and quadratic for second-order).
To obtain the quadratic convergence expected from the \gls{pde} discretization, a second-order scheme is required.
This further motivates the necessity to support higher-order interpolation schemes in communication routines although this usually entails additional implementation effort.
\begin{figure}
    {
        \setlength{\tabcolsep}{4pt}
        \renewcommand{\arraystretch}{1.025}
        \begin{tabular}{|c|c|c|c|c|c|c|}
            \hline
            \makecell{Block \\ Size} & \multicolumn{2}{c|}{\makecell{Constant-Order \\ Scheme}} & \multicolumn{2}{c|}{\makecell{First-Order \\ Scheme}} & \multicolumn{2}{c|}{\makecell{Second-Order \\ Scheme}} \\
            \hline
            & \makecell{$L_2$ Error \\ Norm} & \makecell{$\kappa$} & \makecell{$L_2$ Error \\ Norm} & \makecell{$\kappa$} & \makecell{$L_2$ Error \\ Norm} & \makecell{$\kappa$} \\
            \hline
            $16^3$  & 0.235 & $-$   & 1.067e-2 & $-$   & 3.017e-3 & $-$ \\
            $32^3$  & 0.227 & 0.966 & 5.255e-3 & 0.493 & 7.215e-4 & 0.239 \\
            $64^3$  & 0.224 & 0.984 & 2.611e-3 & 0.497 & 1.764e-4 & 0.244 \\
            $128^3$ & 0.222 & 0.992 & 1.302e-3 & 0.499 & 4.361e-5 & 0.247 \\
            $256^3$ & 0.221 & 0.996 & 6.499e-4 & 0.499 & 1.084e-5 & 0.249 \\
            \hline
        \end{tabular}
    }
    \caption{$L_2$ error norm experiments over multiple resolutions with different interpolation orders.}
    \label{table:err_analysis}
\end{figure}

\section{Benchmark Results}\label{sec:bench}

\par
To highlight the effectiveness of our refinement approach, we perform benchmarks for the Poisson application on the SuperMUC-NG CPU cluster, focusing on weak scaling results.
We compare the performance of our generated communication scheme with the manual baseline implementation, utilizing code generation for all other multigrid kernels with automatic specializations for local refinement.
The cluster consists of Skylake Intel Xeon Platinum 8174\;processors with 24 physical cores.
Each node is equipped with two sockets and 96 Gigabytes of main memory.
We use MPI as our parallelization back-end and employ a 1:1 mapping between an MPI process and a physical core.
The generated code from ExaStencils is automatically vectorized using AVX512 intrinsics.
\par
While load imbalances are omnipresent in such application scenarios due to the varying workload in the boundary handling and the (un-)packing kernels,
we adapted the refinement selection in our benchmarks to mitigate imbalances originating from a different amount of blocks per MPI process.
The experiments capture the transition from using one socket to one or multiple nodes.
The domain $\Omega \in [0, 1]^3$ is partitioned into $N_{cores}$ root blocks which are first globally refined once.
A quarter of the $8 * N_{cores}$ blocks is then locally refined once.
All blocks are then distributed such that each MPI process holds a total of 22 blocks throughout all measurements.
A block comprises $64^3$ cells on the finest multigrid level and $4^3$ cells on the coarsest multigrid level.
In each benchmark, we execute 100 $V(3,3)$ multigrid cycles with a fixed number of 256 coarsest-grid iterations and measure the average time $t_{cycle}$.
\par
\begin{figure}
    \begin{adjustbox}{minipage=\linewidth, scale=1.0}
        \begin{subfigure}{0.5\linewidth}%
            \centering
            \includegraphics[width=1.0\linewidth]{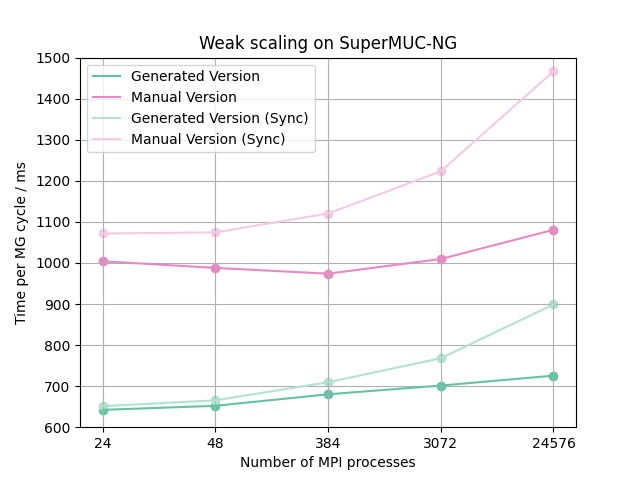}
            \caption{Comparison of $t_{cycle}$ with different timing strategies.}
            \label{subfig:weak_scaling_comparison}
        \end{subfigure}
        \hfill%
        \begin{subfigure}{0.5\linewidth}%
            \centering
            \includegraphics[width=1.0\linewidth]{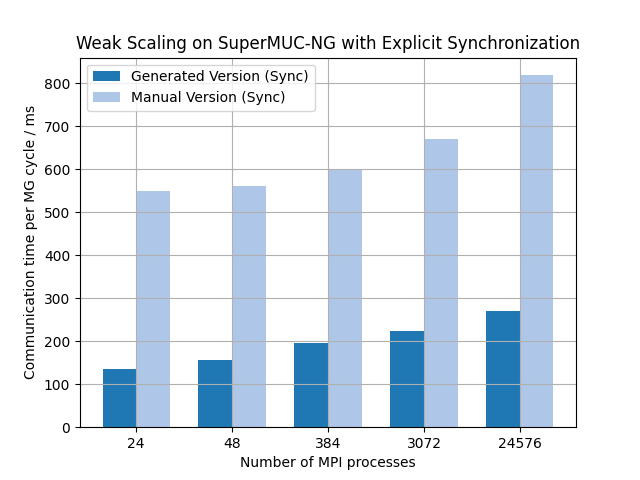}
            \caption{Comparison of communication time in $t_{cycle}$ with synchronized timers.}
            \label{subfig:weak_scaling_stacked}
        \end{subfigure}
    \end{adjustbox}
    \caption{Weak scaling results on SuperMUC-NG with up to 512 nodes.}
    \label{fig:weak_scaling}
\end{figure}

The weak scaling results for both implementations are depicted in~\cref{fig:weak_scaling}.
\Cref{subfig:weak_scaling_comparison} shows the comparison between the two communication implementations with two different timing strategies,
namely with and without explicit MPI synchronization before and after each timing.
These are denoted by lighter and darker colors, respectively.
\par
At first, we will focus on the measurements without synchronization.
It can be observed that both implementations scale well, even for large node counts.
Across all scales, the version with a generated communication achieves a speedup of at least 1.43x.
To assess the performance of our approach, we have set up a light-speed model for the 7-point Jacobi smoother on the finest multigrid level on a single node including communication.
For one lattice site update (LUP) within the kernel, we have one store and seven load operations for the solution and one load for the right-hand side.
Assuming caching and non-temporal stores of the solution vector, we obtain a memory throughput of $n_b = (1 + 1) * 8B + 1 * 8B = 24B$ per cell.
We employ the STREAM benchmark to obtain the effective bandwidth per node and measure $b_s = 154.9$ GB/s.
With that, the maximum expected performance is $P_{max}^{node} = \frac{b_s}{n_b} = 6.45$ GLUP/s per node,
or assuming an equal distribution of the bandwidth across between cores, $P_{max}^{core} = 134.375$ MLUP/s per core.
Throughout all scales, the generated version achieves $98.86 - 102.24$ MLUP/s per core, which equals at least 73.5\% of our maximum expectation.
The manual version, on the other hand, achieves $87.39 - 98.20$ MLUP/s per core, which results in a minimum of 65\% of our expectation.
While this is already a significant improvement on the finest multigrid level,
the speedup factor of $t_{cycle}$ in the comparison also comes from avoiding overheads in the coarser-grid communication with domain-specific optimizations such as the proposed pointer prefetching.
On the coarsest grid, the generated version is $7.90 - 19.21$ faster as the waLBerla communication was not designed for this special case.
The runtime fraction of the coarse-grid solver with generated communication is $1.64 - 5.15\%$ and with the manual version, it is $20.17 - 27.35\%$.
\par
As communication is our main optimization target, we further want to quantify the improvements to the communication time to explain the scaling behavior.
Thus, explicit synchronization was used to avoid capturing imbalances in the communication timings.
The communication time spent in a cycle for both implementations is depicted in~\cref{subfig:weak_scaling_stacked}.
As expected, both approaches show only minor runtime increases when transitioning from one socket to one node
and have greater increases between one and multiple nodes distributed on several islands.
We observe that the generated communication roughly achieves a three- up to fourfold speedup across all measurements,
reducing the communication fraction in the rough range of $51-56\%$ to $20-30\%$.
Note that $t_{cycle}$ also captures synchronization time and, thus, the speedup does not directly translate over to the communication fraction ranges.
\section{Conclusions}\label{sec:conclusion}
\par
In this work, we bridge the gap between code generation technology for multigrid solvers and block-structured meshes with local refinement.
We propose data exchange routines between coarse and fine blocks augmented with interpolation schemes that maintain error convergence despite local jumps in the mesh resolution.
These interpolation schemes do not rely on values from neighboring blocks and minimize the volume of data that needs to be communicated to efficiently use highly parallel hardware.
We incorporate mesh refinement concepts into our code generator to generate specialized kernels and communication routines for multigrid solvers with local refinement.
We also investigate the impact of the order of an interpolation scheme on the error convergence for a locally refined 3D Poisson equation.
The benefits of generating the communication routines are demonstrated in a comparative study with scaling results on the SuperMUC-NG CPU cluster using up to 512 nodes.
\section{Future Work}\label{sec:future}
\par
This work presents a software infrastructure for code generation of multigrid on octrees which can be used as a foundation for multiple avenues.
\par
This includes employing our approach for more complex application scenarios such as \glspl{pde} with variable coefficients or for real-world scenarios such as concentration transports within a fluid.
Depending on the stencil, these applications could require diagonal elements to be communicated which needs extensions to the existing interpolation schemes.
\par
Another avenue is extending the mesh refinement implementation in the code generator with new features.
This includes the support for other variable localizations than cell-centering.
Optimizing the packing of our C2F schemes to employ a ring-buffer data structure for reusing intermediate extrapolation results for neighboring cells would decrease the time spent in communication even further.
This can be done by writing code transformations to prepare the loop nests for the interpolation for the loop-carried \gls{cse} optimization from~\cite{Kronawitter2016}.
Achieving performance portability by supporting GPU platforms is another important milestone in the future and calls for platform-dependent optimizations such as using shared memory for storing reusable intermediate results.

\section*{Acknowledgment}
This work is supported by the German Research Foundation (DFG) as part of the program "Dynamic HPC software frameworks: Seamless integration of existing simulation frameworks and code generation techniques" under grant number KO 46641/3-1.
We also acknowledge the Leibniz Supercomputing Centre for providing access to the SuperMUC-NG computing resources.
We thankfully acknowledge the scientific support and HPC resources provided by the Erlangen National High Performance Computing Center (NHR@FAU) of the Friedrich-Alexander-Universit{\"a}t Erlangen-N{\"u}rnberg (FAU). The hardware is funded by the DFG.

\printbibliography{}

\end{document}